\documentclass{elsart}

\usepackage{color}
\usepackage{epsfig}
\usepackage{amssymb}
\usepackage{graphicx}

\begin{document}
\begin{frontmatter}
\title{Heavy Nuclei Enrichment of the Galactic Cosmic Rays at 
High Energy: Astrophysical Interpretation}
\author{David Maurin},
\address{LAPTH, Chemin de Bellevue, BP 110, 74941 Annecy le Vieux 
CEDEX, France
\\ IAP, 98 bis Bd Arago, 75014 Paris, France}
\author{Michel Cass\'e},
\address{SAp, CEA, Orme des Merisiers, 91191 Gif/Yvette CEDEX, 
France,\\
 IAP, 98 bis Bd Arago 75014 Paris, France}
\author{Elisabeth Vangioni-Flam},
\address{IAP, 98 bis Bd Arago 75014 Paris, France}
\thanks[CorrAuth]{Corresponding author.\\
{\em E-mail address:} maurin@lapp.in2p3.fr (D. Maurin)}



\begin{abstract}
A substantial increase of the mean logarithmic mass $\langle 
\ln A\rangle$ 
of galactic cosmic rays {\em vs} energy
 has been observed . We study three effects that could explain this 
 trend i) different source spectra for protons and heavy nuclei ii) 
 a selective nuclear destruction in flight of heavies iii) a gradient
 of the source number and chemical composition in the galactic disk.
 We take advantage of the diffusive cosmic ray propagation model 
 developed at LAPTH to study specifically the geometrical aspects 
 of the propagation and extend it to high energy. Using a simple 
 modeling of the spectral knee around $10^{15}$~eV, a bump in 
 $\langle \ln A\rangle$ appears. This feature is smoother when the 
 spectral index of protons is steeper than Fe's.
 We analyze the effects of the rigidity dependence of the diffusion 
 coefficient and the scale height of the confinement halo and we 
 show that $\langle \ln A\rangle$ is most sensitive to the first 
 parameter. Pure geometrical effects are less determining than the 
 diffusion coefficient spectral index. Subsequently, we conclude 
 that the physics of cosmic ray confinement is the essential cause 
 of the heavy nuclei enrichment until $\sim 10^{15}$~eV.
\end{abstract}
\end{frontmatter}

\section*{Introduction}

Recent measurements of the cosmic ray average logarithmic mass and 
all-particle spectrum  around $10^{15}$~eV~\cite{Apanasenko,Shirasaki} 
give new clues to understand the origin of the cosmic rays
 and in particular the puzzle of the knee in the energy spectrum.
Analysis of such data in a coherent theoretical framework is a rough 
task, and even if much progress has been done both in theoretical and 
experimental sides, none of the models proposed so far to solve this 
problem has been unanimously accepted. The highest energy particles 
are almost certainly extragalactic. A similar origin is not excluded 
near the knee, but it is difficult to account for the observed 
continuity of the spectrum in this region~\cite{Axford}. As a 
consequence, the intermediate region between $10^{15}$ and
 $10^{19}$~eV should be analysed in terms of the same physical 
 mechanisms than lower energy particles.

There are at least three explanations for the knee: (i) a change 
in propagation parameters (diffusive regime),
 (ii) a change in 
the source regime, (iii) a change related to the properties of high energy
interactions in the atmosphere or a subtle combination of all three.
As Schatz~\cite{Schatz} emphasized recently, the fine structure of the knee
in all-particle spectra provided by extensive air showers can help to 
discriminate between these solutions.
Furthermore, recent data from collider 
seem to show no drastic 
departure from cross sections predictions in the range 
100~--~1000~TeV~\cite{Apanasenko}. Anyway, this paper will only 
concentrate on astrophysical aspects.

More information can be obtained by measuring $\langle \ln A\rangle$.
All-particle spectrum and $\langle \ln A\rangle$ are given by
linear combination of the individual fluxes with different weights;
\begin{displaymath}
\Phi^{\rm all} \equiv \sum_j \Phi_j
\end{displaymath}
and
\begin{displaymath}
\langle \ln A \rangle \equiv \frac{\sum_j \ln A_j \Phi_j}{\sum_j \Phi_j}\;.
\end{displaymath}
Therefore they provide different information of these fluxes.
Obviously, these weighted quantities are not 
very useful at ``low energy" where all nuclei are well resolved in 
satellite or balloon experiments (for a compilation of data, 
see~\cite{Wiebel-SoothII}). 
Experimental difficulties arise around and above the knee: fluxes are 
very low ($\sim$ m$^{-2}$ sr$^{-1}$ yr$^{-1}$) and large ground array 
detectors are needed to collect unresolved events with a good statistic.
Incidently, the all-particle spectrum can be extracted {\em via} shower 
parameters ({\em e.g.} core position, direction,...) and an estimation of 
the average logarithmic mass is also possible through various 
methods ({\em e.g}~\cite{Shirasaki,Bernlohr,Swordy} and in 
particular~\cite{Antoni1}).
A new experimental technique has been recently proposed, which can 
potentially yield excellent charge resolution measurements near and above 
the knee~\cite{Kieda}. But at the present time, even if a few solutions 
exist to infer the composition~\cite{Roberts,Linder},  all-particle flux, 
mean logarithmic mass and sometimes proton and helium 
spectra~\cite{Amenomori3} are grossly the only available observables 
near and above the knee (for a detailed revue, see~\cite{Swordy3}).

The paper is organized as follows: (i) features of
primary species propagated in diffusion model are reminded; (ii)
simple models are used to explore the behaviour of $\langle \ln A\rangle$ 
considering separately the effect of three parameters (source spectra, 
propagation, geometrical aspects);
 (iii) the results of these simplified models are analysed;
 (iv) using a more realistic simulation, the three parameters 
are discussed in details, 
and (v) all these parameters are included in the propagation model of 
Maurin et al.~\cite{PaperI} and are finally combined to analyze the 
knee problem.

\section{Basic features of propagation models}

For practical reasons, spectra at low energy are almost always displayed
in units of kinetic energy per nucleon, because 
this quantity is conserved in nuclear reactions. This
convention has become the rule in the analysis of cosmic ray propagation,
 {\em e.g.} secondary to primary ratio studies.
 Above PeV energies, fluxes are plotted {\em vs} energy per particle 
because it is the only observable provided by ground observatories. 
Whereas various primary species present similar spectra when plotted 
versus kinetic energy per nucleon or rigidity, they behave differently 
in terms of total energy (compare for example Fig.~5-a of~\cite{Simpson} 
that displays fluxes in units of kinetic energy per nucleon  
and Fig.~2 of~\cite{Wiebel-SoothII} that shows the same fluxes, 
but in units of total energy); so a combination of all the corresponding 
spectra into a single quantity (for example the average logarithmic mass) 
will provide a different dependence in terms of total energy or in terms 
of rigidity.

\subsection{Propagation models above the knee and extrapolation to higher
energies}
\label{musea}
Although cosmic ray properties are well understood
up to a few hundreds of GeV/nuc (see the recent review of~\cite{Webber}), 
it is difficult to harmonize all the observables ({\em i.e.} protons, 
nuclei, e$^-$, e$^+$, $\bar{p}$, $\gamma$ rays; see for 
instance~\cite{Strong}). In particular, the hypothesis according to 
which the fluxes measured locally are representative of the fluxes 
present everywhere in the Galaxy is still debated.
At intermediate energies, extension of the usual propagation models
has mainly to face the problem 
that the observed anisotropy is
very low. This seems to favor a Kolmogorov spectrum (interstellar
 turbulence) associated to the rigidity dependence
 of the diffusion coefficient (see discussion in~\cite{Jones}).

Cosmic ray spectra are affected by the propagation process.
Maurin et al.~\cite{PaperI} have recently developed a propagation model     
(see Sec.~\ref{Propagation_prescription} for details)
that is in principle valid 
for a wide energetic range as long as charged nuclei are considered.
This is a two zones model (thin gazeous disc -- half-height
$h=100$ pc -- and large diffusive halo $L\sim3-10$ kpc) 
where cylindrical symmetry is assumed (radial extension, $R=20$ kpc).
The steady--state differential density $N^j(E,\vec{r})$ of the nucleus $j$ as a
function of energy $E$ and position $(r,z)$ in this model
is given by
\begin{eqnarray}
      {L}_{dif\!f}N_{j}(r,z)+ 2h\delta(z) \left(
         q^j_0Q_j(E)q(r)+\sum_{k=1}^{j-1}\Gamma^{kj}N_{k}(r,0)
         -\Gamma^j N_{j}(r,0)\right)\nonumber\\
        =2h\delta(z) \frac{\partial}{\partial E}
        \left\{ b^j(E)N_j(0) - d^j(E)\frac{\partial}{\partial E}
       N_j(0)\right\}\nonumber
\end{eqnarray}      
\begin{eqnarray}
           {L}_{dif\!f}\equiv-V_{c} \frac{\partial}{\partial z}
          +K(E)\left(\frac{\partial^{2}}{\partial z^{2}}+
          \frac{1}{r}\frac{\partial}{\partial r}
        (r\frac{\partial}{\partial r})\right)\nonumber
\end{eqnarray}
The operator ${L}_{dif\!f}$ represents convection  ($V_c$) plus spatial 
diffusion ($K(E) = K_0 \, \beta \times R^{\delta}$) acting in the whole 
box. All other terms describe processes localized
in the thin gazeous disc only: the bracket corresponds to primary source term,
secondary spallative production from heavier nuclei, and destruction 
cross section (radioactive-induced processes have been omitted here). 
Curly bracket provides all terms leading to energetic redistribution, {\em i.e}
losses (Coulombian, ionisation and adiabatic expension losses) and 
gains (reacceleration) described by two effective parameters $b_j(E)$ 
and $d^j(E)$ (for all details, see~\cite{PaperI}).

In the leaky box model, widely used, all the quantities are
spatially averaged. The 
diffusion-convection term is then replaced by 
an effective escape term which has the meaning of a residence time $\tau_{esc}$
($\sim 20$ Myr at a few GeV/nuc) in the confinement volume ($-L_{dif\!f}
\leftrightarrow \tau_{esc}^{-1}$).
If energy gains and losses are discarded, the leaky box equation reads
\[
-\frac{\bar{N}_j}{\tau_{esc}}+\left(\bar{q}^j_0Q_j(E)+
\sum_{k=1}^{j-1}\Gamma^{kj}\bar{N}_{k} -\Gamma^j \bar{N}_{j}\right)=0
\]
As an immediate consequence, leaky box models do not allow to take in 
consideration, for example, a radial dependence of Galactic sources.
More generally, any subtle effect related with spatial dependence 
of any terms of the diffusion-convection equation is automatically swept away.

\begin{figure}[ht]
\includegraphics*[width=0.9\textwidth]{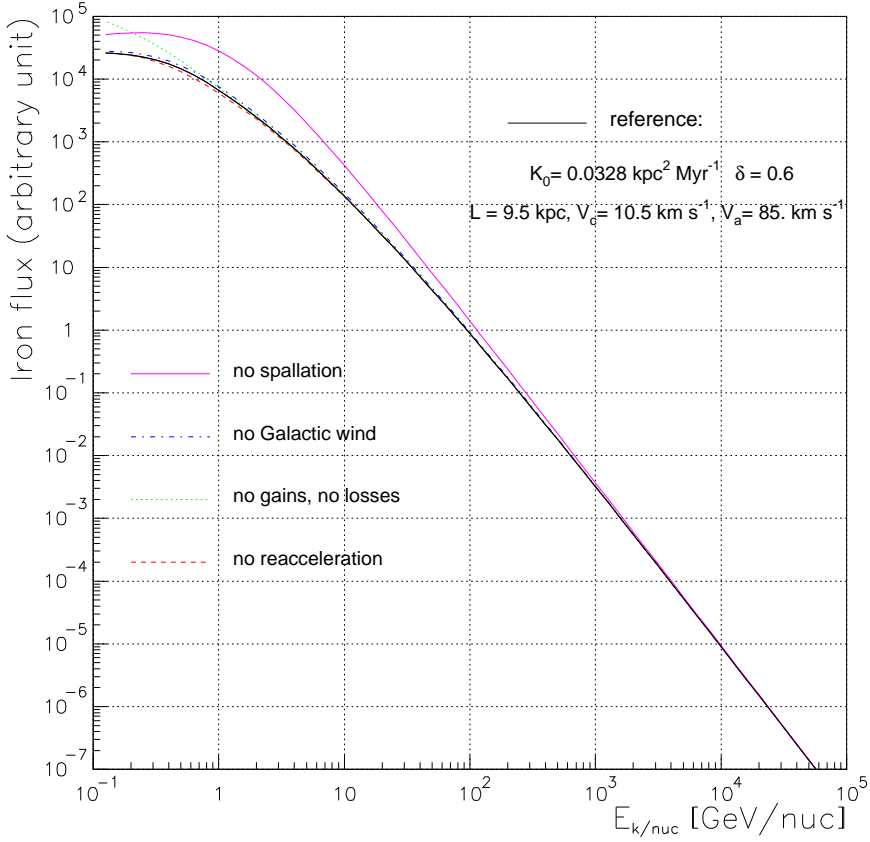}
\caption{Unnormalized Fe flux for a typical propagation model 
that reproduces B/C (reference curve); the five parameters (diffusion~-- 
normalization $K_0$ and slope $\delta$~--, 
halo size $L$, convection $V_c$, reacceleration $V_a$) are taken
from~\cite{PaperI}. Other curves correspond to the same 
propagation parameters, except one that has been switched off.
}
\label{Fe}
\end{figure}
Figure ~\ref{Fe} displays the iron flux as a function of kinetic
energy per nucleon in our diffusion-convection model; 
this behaviour is standard for all charged nuclei except protons, 
for which propagation is less sensitive to nuclear reactions. 
Within realistic propagation 
models, several parameters affect the primeval power law source spectrum:
\begin{itemize}
  \item The most spectacular effect is that of inelastic interactions 
  (nuclear destruction). It becomes negligible ($\lesssim$ a few \%) 
  between 1~TeV/nuc and 10~TeV/nuc,
  depending on the species, since the escape time is shorter than
 the nuclear time.
  \item The spectrum is also affected by energy losses up to $\sim 
  100$~GeV/nuc.
  \item More controversial effects may be present:
  galactic wind up to 1~TeV/nuc and reacceleration up to some tens of 
  GeV/nuc.
\end{itemize} 
Propagation effects are present up to 10~TeV/nuc. This induces
modifications of spectra  up of a few hundreds TeV in total energy,
at least for heavier nuclei ({\em i.e.} Fe) leading in turn to a
dependence on energy of the average logarithmic mass till these higher 
energies, {\em i.e.} near the knee ($\sim$~PeV). Since current high energy
data are provided in total energy per particle, all results will be given 
in total energy (denoted $E$) unless stated otherwise.
\section{Separation of key ingredients}
\label{key}
Solar modulation is ignored because it has almost no effect on fluxes 
above a few tens of GeV/nuc. Three items (source 
spectra, propagation and geometrical effects) are pieces of 
the cosmic ray puzzle.
In this section, we introduce simplified models where only two species 
are considered ({\em e.g.} H and Fe) that allow to analyze qualitatively 
the increase of the mean mass of cosmic rays with energy.
  \subsection{Source spectrum effect}
\label{pure source}
In the simplest propagation model that one can imagine, pure diffusion
is considered and only two 
ingredients are necessary:
\begin{equation}
Q_j(E)=q^0_j\times \left(\frac{R}{1 \mbox{~GV}}\right)^{-\alpha_j}
 \mbox{ is the source spectrum,}
\label{sourceQ}
\end{equation}
\begin{equation}
 K(E)=K_0\times \left(\frac{R}{1 \mbox{~GV}}\right)^{\delta}
 \mbox{ is the diffusion coefficient.}
\label{diffusionK}
\end{equation}
In the above expressions, $R=p/Z$ is the rigidity,
$q^0_j$ and $\alpha_j$ are respectively the source abundance and
the spectral index of nucleus $j$,
$K_0$ is the normalization coefficient of the diffusion coefficient
and $\delta$ its slope (both are assumed independent of the species 
considered). The detected local flux (in arbitrary units) is thus 
expressed merely as the ratio of sources over the diffusion coefficient 
(up to a coefficient $v/4 \pi\simeq c/4\pi$ for all nuclei):
\begin{equation}
\Phi_j(E)=Q_j(E)/K(E).
\end{equation}
This model can be viewed as the high energy 
limit of all diffusion models. This limit is also that of
a locally equivalent leaky box model, where once emitted,
the nucleus can escape from the confinement box with a
probability which is related very simply to $K(E)$~-- see for 
example~\cite{PaperI} and references therein.

\subsubsection{$\langle \ln A\rangle$ for the two nuclear component model}
\label{trucmuche}
In a two nuclei model, the evaluation
of the Fe/p ratio and the average logarithmic mass number~-- 
``average mass" for short~-- is straightforward.
At a given energy, the rigidities of two charged nuclei are 
different. Indeed,
the term $K(E)$ can't be simply factorized, and some residual
multiplicative constant depending on $\delta$ appears. Using
the approximate relation  $R=p/Z\simeq E/Z$ (E is the total energy 
per particle) which is correct beyond a few tens of A~GeV,
 we have
\begin{displaymath}
    {\rm Fe}/{\rm p} (E)=\left(\frac{q^0_{\rm Fe}}{q^0_{\rm p}}
    \times
    \frac{Z_{\rm Fe}^{\alpha_{\rm Fe}+\delta}}{Z_{\rm p}^{\alpha_{\rm 
    p}+\delta}}\right)_{100\rm~GeV} 
    \times
    \left(\frac{E}{100 {\rm~GeV}}\right)^{-\alpha_{\rm Fe}+\alpha_{\rm p}}\:,
    \end{displaymath}
where the first term corresponds to the relative source abundance Fe/p
taken at the energy 100~GeV, which is $\sim 1/20$ (see Fig.~2 of
compilation of~\cite{Wiebel-SoothII}).
Apart from numerical constant, we roughly see that the steeper 
${-\alpha_{\rm Fe}+\alpha_{\rm p}}$ effective slope, the greater will 
be the average mass. To be more precise,
\begin{equation}
        \langle\ln A \rangle (E)= \frac{\ln (A_{\rm Fe})}{\left(1+ 
        \left[ (q^0_{\rm p}/q^0_{\rm Fe})
        (Z_{\rm p}^{\alpha_{\rm p}+\delta}/Z_{\rm Fe}^{\alpha_{\rm Fe}
        +\delta})\right]_{\rm 100~GeV}
        \times (E/100 {\rm~GeV})^{-\alpha_{\rm p}+\alpha_{\rm Fe}}\right)}
\label{Pure_index}
\end{equation}
whereas in terms of the all-particle spectrum, the evolution~-- up to a 
normalization~-- can be written as
\begin{equation}
        \Phi^{\rm all}\propto E^{-\alpha_{\rm Fe}-\delta}\times
        \left(1+ 
        \left[ \frac{q^0_{\rm p} Z_{\rm p}^{\alpha_{\rm p}+
        \delta}}{q^0_{\rm Fe}Z_{\rm Fe}^{\alpha_{\rm Fe}+\delta}}
        \right]_{\rm 100~GeV}
        \times \left(\frac{E}{100 {\rm~GeV}}\right)^{-\alpha_{\rm p}+
        \alpha_{\rm Fe}}\right)\;.
\label{Pure_index_all_particle}
\end{equation}
We remark that in our simple two species model, $\Phi^{\rm all}\propto 
E^{-\alpha_{\rm Fe}-\delta}/\langle\ln A \rangle (E)$; the behaviour 
of all-particles spectrum will be discussed in Sec.~\ref{gfdsa}. 

\subsubsection{Application to the``wind/ISM" supernova model}

\label{windISM}
Among the various acceleration models, some of them
are able to produce various spectra for different
nuclei~(see Sec.~\ref{biermanie}).
The maximal source effect is obviously obtained when $|-\alpha_{\rm p}+
\alpha_{\rm Fe}|$ is maximal. 
For typical values $-\alpha_{\rm p}+\alpha_{\rm Fe}
\sim -0.1$ such as those advocated by the wind/ISM supernovae model of 
Biermann and collaborators~\cite{Wiebel-SoothI}, and with the above value 
$\left[ (q^0_{\rm p}/q^0_{\rm Fe}) (Z_{\rm p}^{\alpha_{\rm p}+
\delta}/Z_{\rm Fe}^{\alpha_{\rm Fe}+\delta})\right]_{\rm 100 GeV}\sim 20$,
we obtain
\begin{equation}
 \langle\ln A \rangle (E={\rm 100\:GeV /\: 10\:TeV /\: 1\:PeV/\:
 \infty)}\sim 
 0.19\: / \: 0.30 \: / \:0.50\: / \:4.03\\
\label{result1}
\end{equation}
Indeed, the evolution is expected to be smoother
 when all cosmic ray nuclei are considered, because the slopes of nuclei 
heavier than hydrogen are similar (see Sec.~\ref{biermanie}); 
but this sole effect leads to an evolution of the average mass.
Incidently, if $\alpha_{\rm p}=\alpha_{\rm Fe}$, we are left with a 
constant $\langle\ln A \rangle$ that merely depends on the relative 
abundances of nuclei. This conclusion holds again when several nuclei 
are considered.

  \subsection{Propagation effect: same spectral index}
\label{Propag_Effect}

The process of inelastic collisions -- which differs from 
one nucleus to another -- has a major impact on propagation (see Fig.~\ref{Fe}).
 To obtain order of magnitude
estimates, a simple leaky box description including just destruction cross
sections is sufficient.
Thus, the propagated flux for a primary species $j$ is:
\begin{equation}
\Phi_j(E)=\frac{Q_j(E)}{1/\lambda_{esc}(E)+\sigma^{inel}_j/{\bar A}}
=\frac{1}{\bar{A}}\times\frac{Q_j(E)}{\sigma_{esc}(E)+\sigma^{inel}_j}
\label{LB}
\end{equation}
Here $\sigma^{inel}_j=\sigma^{tot}_i-\sigma^{el}_i$ is the total 
inelastic (or reaction) cross section of species $j$, and 
$\lambda_{esc} (E)$ is the usual escape length of the leaky 
box in g cm$^{-2}$ and $\bar{A}$ is the mean mass of the atoms in 
the interstellar medium.
This can also be rewritten in terms of an effective escape cross section
$\sigma_{esc}$ (in cm$^2$) that is related to the usual grammage by a 
simple factor.

Under the assumption of the same source
spectrum ({\em i.e.} universal slope $\alpha_j=\alpha$ for all species)
and since there is no geometry in the leaky box, assuming
an escape dependence $\lambda_{esc}(E)=\lambda_0 \beta
R^{-\delta}\simeq \lambda_0 R^{-\delta}$, {\em i.e}  $\sigma_{esc}(E)
=\sigma_0 R^{\delta}$ for all nuclei, we  get:
       
\begin{equation}
        \langle\ln A \rangle (E)=\ln (A_{\rm Fe})\times\left[
        \left(1+ 
        (q^0_{\rm p}/q^0_{\rm Fe}) \times
         \left(\frac{Z_{\rm p}}{Z_{\rm Fe}}\right)^{\alpha}    
        \times
        \frac{(\sigma_0(E/Z_{\rm Fe})^{\delta}+
    \sigma^{inel}_{\rm Fe})}{(\sigma_0(E/Z_{\rm p})^{\delta}+
    \sigma^{inel}_{\rm p})}\right)\right]^{-1}
\label{Pure_propagation}
\end{equation}

Contrary to the previous case, the ratio $q^0_{\rm p}/q^0_{\rm Fe}$ 
does not depend on energy.

\subsubsection{Leaky box results}
\label{verte}
The typical diffusion coefficient for leaky box models is given for
example by~\cite{Webber2}
\begin{displaymath}
\lambda=35.1\beta\:(R/1 {\rm~GV})^{-0.61} \mbox{~g cm}^{-2}
\end{displaymath}
which corresponds in the above expression to $\sigma_0\sim 47$ mb.
We have in this model $\alpha\sim 2.2$, 
$\sigma^{inel}_{\rm p}\sim 30$ mb (Particle Data 
Group\footnote{http://pdg.lbl.gov/}),
$\sigma^{inel}_{\rm Fe}\sim 710$ mb~\cite{Carlson}.
In order to compare this effect with the precedent, we set 
$q^0_{\rm p}/q^0_{\rm Fe}$ to $\sim 25000$ which gives
the same $\langle\ln A \rangle$ at 100~GeV as before. Thus,
\begin{equation}
 \langle\ln A \rangle (E={\rm 100\:GeV /\: 10\:TeV /\: 1\:PeV \:/\: \infty)}
 \sim 0.19\: / \:0.83 \: / \:1.05 \: / \:1.08
\label{result2}
\end{equation}

Even near the knee energies, the average mass slightly evolves due to 
spallations.
It is then clear from formula~(\ref{Pure_propagation}) that the evolution 
of chemical abundances depends on (i) the destruction cross section, 
(ii) the diffusion coefficient slope $\delta$, as well as other
propagation parameters (see Sec.~\ref{musea}).
  \subsection{Geometrical effects}
  \label{metalicite}
In homogeneous diffusion models ({\em e.g.} leaky box models), 
this would be the final step.
In the context of diffusion models,
 the nuclei propagate in a two zones/three dimensional space,
 and we take into account the gradients of the source number and of the
 metallicity.

        \subsubsection{Basic description}
\label{greswc}
 Let first assume that spectral source indexes are similar for 
 all accelerated species, and that pure diffusion prevails. The 
 quantity
\begin{equation}
        \langle\ln A \rangle (E)= \frac{\ln (A_{\rm Fe})}{1+ 
        \left(q_{\rm p}/q_{\rm Fe}\right)^{\rm eff}}\;,
\label{Pure_gradient}
\end{equation}
depends on energy {\em via} the relative abundance 
$(q_{\rm p}/q_{\rm Fe})^{\rm eff}\equiv (q^0_{\rm p}/q^0_{\rm Fe})$.
Actually, the source term depends on radial coordinate, and the average
distance $\langle r_{\rm X}\rangle$ from which a given species $X$ come
depends 
on energy. At a given energy, $\langle r_{\rm X}\rangle$ is generally 
different for two species, implying an indirect dependence of 
$(q_{\rm p}/q_{\rm Fe})^{\rm eff}\equiv (q^0_{\rm p}/q^0_{\rm Fe})$ on
energy. Moreover, even if these average distances $\langle 
r_{\rm p}\rangle (E)$
and $\langle r_{\rm Fe}\rangle(E)$ are equal, the relative abundances may 
depend on this distance; this is the {\em pure} metallicity effect. 
At sufficiently high energy, {\em i.e.} 
a few tens of TeV/nuc, $\langle r_{\rm X}\rangle$ is a constant
number that only depends on the size of the diffusive box: no 
geometrical effects are expected.

        \subsubsection{Metallicity effect}
We focus on  the {\em pure} metallicity effect: forgetting for a while 
the radial source distribution effect, 
we suppose that relative distribution of species only depends of their 
location.
Namely, we take a gradient
$\nabla [{\rm Fe}/{\rm H}]\sim -0.05$ dex kpc$^{-1}$, corresponding to an
 increased metallicity towards the center of the Galaxy (see 
 Sec.~\ref{gradientmetal}).
The sun is located 
at $r=R_{\odot}$ (8.5 kpc). The metallicity gradient is given by:
\begin{displaymath}
[{\rm Fe}/{\rm H}]=-0.05\: (r-R_{\odot}) \equiv
\log \left(q_{\rm Fe}/q_{\rm p}\right)^{\rm eff}
-\log ({\rm Fe}/{\rm H})_{\odot}
\end{displaymath}
leading to
\begin{equation}
\left(\frac{q_{\rm p}}{q_{\rm Fe}}\right)^{\rm eff}=
\left(\frac{q_{\rm p}}{q_{\rm Fe}}\right)_{\odot}\times
10^{0.05\:(r-R_\odot)}\:.
\label{qwerty}
\end{equation}
A rough estimation of the $r$ dependent term above can be obtained
as follows: at high energy, cosmic ray nuclei
cannot come farther than the center, {\em i.e.} $r=0$
kpc. At lower energy (say 100~GeV), they all come from
$r=R_\odot$. It corresponds to
\begin{displaymath}
\left(\frac{q_{\rm p}}{q_{\rm Fe}}\right)_{\rm 1\: PeV}\sim
0.32\left(\frac{q_{\rm p}}{q_{\rm Fe}}\right)_{\rm 100\:GeV}
\end{displaymath}

This crude evaluation that overestimates some effects gives
us the evolution of the mean mass for the metallicity gradient effect: 
using Eq.~(\ref{Pure_gradient})
and assuming once again that at 100~GeV, the ratio
$\left(q_{\rm p}/q_{\rm Fe}\right)^{\rm eff}\sim 20$, we obtain
\begin{equation}
 \langle\ln A \rangle (E={\rm 100\:GeV /\:\infty)}
 \sim 0.19\: / \: 0.54
\label{result3}
\end{equation}
 This {\em pure} metallicity effect is small compared to the others.
 Moreover, as cosmic ray sources are distributed in the
anticenter (lower metallicity) as well as in the center (higher metallicity),
we expect a sort of cancelation when both contributions are added.
The same balance is likely to occur for the second geometrical effect.

\section{First conclusions}
\label{gfdsa}
\begin{figure}[ht]
\includegraphics*[width=\textwidth]{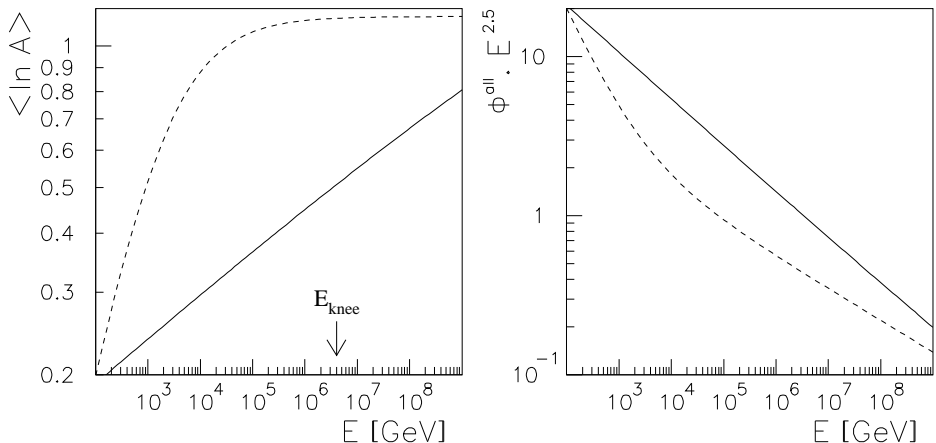}
\caption{Two nuclei model, p and Fe. Left panel: $\langle \ln A \rangle$ 
for {\em pure} spectrum effect~-- solid line (see Sec.~\ref{windISM})  and
{\em pure} propagation effects~-- dashed line (see Sec.~\ref{verte}). 
Right panel: same for all-particle flux (arbitrary units).}
\label{toy_model1}
\end{figure}

Several qualitative remarks can be made about the relative importance 
 of the various effects.  The metallicity gradient plays an almost
negligible role compared to the two others. 
Pure propagation effect is dominant (see Fig~\ref{toy_model1}, 
left panel; right panel shows the two separate effects for the 
all particle flux), 
but as well as geometrical effects it ceases to act around 100~TeV~--~1~PeV 
(roughly the knee energy). Then, the pure source effect only, {\em i.e.} 
different $\alpha$ for protons and other species ($\alpha_{\rm p}\neq
\alpha_{\rm Z>1}$), is able to produce an evolution of the average mass. 
This is an important result that validates 
the approach used in a recent study~\cite{Shirasaki}.

Including all the primaries, we should
expect concerning the first effect -- pure source spectrum, see 
Eq.~(\ref{result1}) -- a smoother evolution, since other primaries 
show slopes similar to that of iron and mix their effects. Concerning 
the second effect, a smoothing is also expected 
since all nuclei are equally dispatched between proton 
and iron (from the cross section point of view).

\begin{figure}[ht]
\includegraphics*[width=\textwidth]{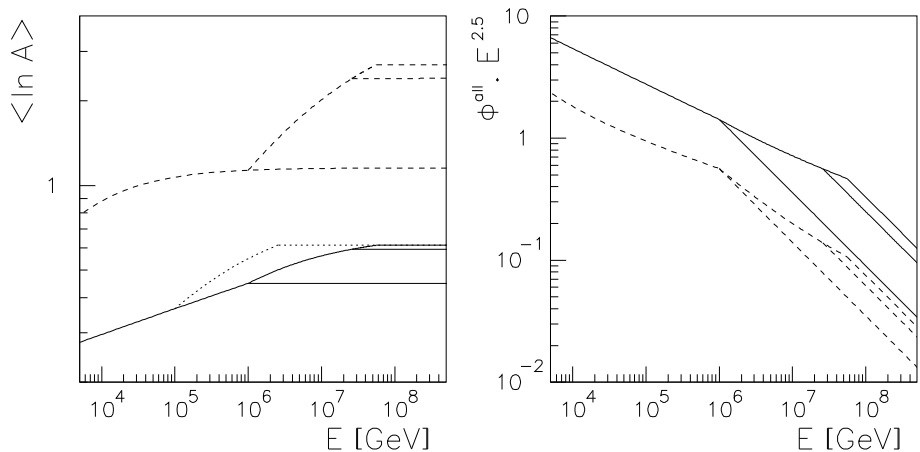}
\caption{Left panel: $\langle \ln A \rangle$, and right panel: all-particle 
flux times $E^{2.5}$ (arbitrary units). The various cases correspond
to the two effects previously studied (see Fig.~\ref{toy_model1})
with an additional modeling for the knee. For each curve (dashed and 
solid line) we have from bottom to top a transition $E_{\rm knee}= 
10^{15}$~eV, 
$E_{\rm knee}= Z\times10^{15}$~eV, and $E_{\rm knee}= A\times10^{15}$~eV.
The dot curve in left panel corresponds to the case
$E_{\rm knee}= Z\times10^{14}$~eV.}
\label{toy_model2}
\end{figure}

 To visualise the changes induced by the knee, we display in 
 Fig.~\ref{toy_model2}
(for the two previous cases) three forms for the break in spectra, 
namely a break at $10^{15}$~eV either in total energy, or rigidity 
or energy per nucleus. Concerning 
$\langle\ln A\rangle$, we see that the evolution from lighter to heavier 
nuclei at the knee is more important if the slope of hydrogen and other 
nuclei are the same below the knee. If not the case, we see in particular
that the farthest the break occurs, the smoother is the bump in 
$\langle\ln A\rangle$ (compare dot line~-- $E_{\rm knee}=10^{14}$~eV~-- 
to solid line~-- $E_{\rm knee}=10^{15}$~eV).
For each model, we have after the knee a constant composition
related to the fact that all nuclei have now the same slope. Concerning
the all-particle spectrum, we see that, except for the situation where 
the break is energy dependent, there is a smooth evolution on about one 
decade before reaching the definitive slope. A first break occurs 
when protons change slope, and the second when Fe does. The
situations where $E_{\rm knee}\propto A$~eV or $E_{\rm knee}\propto 
Z$~eV are quite
similar, and the difference will be rather difficult to see in data.
 Note that a change of diffusive regime (with $\Delta\delta=0.4$) is
completely equivalent to the case $\Delta\alpha=0.4$ in source spectra
if $E_{\rm knee}\propto Z$~eV.
This is because at these energies, subtleties of propagation are 
irrelevant.

\section{Cosmic ray diffusion model}
\label{INPUTS}
   \subsection{Spectral indexes of sources}
   \label{biermanie}

\subsubsection{Acceleration models}
The maximal energy 
reached in shocks associated with supernovae is of about 
$E\sim Z\times 10^{14}$~eV (see for instance~\cite{Lagage}).
As the acceleration processes are rigidity dependent,
there is a cut--off at Z times the maximum energy gained
and these models predict an increase of the average mass of 
primary cosmic rays with total energy. In fact, the limit
$E\sim Z\times 10^{14}$~eV is close to the knee energy
so that another acceleration process must be found for higher 
energies. To bypass
this limit and explain slopes near the knee, several mechanisms 
have been proposed ; for example, acceleration by terminal shock of 
galactic wind~\cite{Jokipii1,Jokipii2}, by neutron star 
quakes~\cite{ChengChi} or by pulsars~\cite{Bednarek}, contribution from
a single recent local supernova explosion~\cite{Erlykin1,Erlykin2},
and photodisintegration of nuclei by a background of optical and soft 
UV photons in the sources~\cite{Candia}. Another
explanation also has been proposed, based on the extension of supernova 
acceleration models, but they are contradictory: it is shown 
in~\cite{Klepach} that acceleration by multiple spherical shocks in 
OB associations (superbubbles)
is not sufficient to reach $E\sim10^{17}-10^{18}$~eV unless extreme 
values of the turbulence parameters are used. A different conclusion is 
drawn in~\cite{Bykov}, where the authors adjust
turbulence parameters to reproduce the data. Recently, going back to the
problem of the maximal energy gained in SN shocks,~\cite{Bell,Lucek} showed 
that non-linear amplifications of the magnetic field by the cosmic rays 
themselves could push the usual limit to $Z\times 10^{17}$~eV, and even 
a factor ten more if stellar wind pre-exists. Note that this distinction 
between {\em standard} supernovae 
and supernova explosions of massive stars in their own wind has been 
advocated by Biermann and collaborators~\cite{Biermann,Biermann2} as 
the possible explanation of the knee. 
This a is very important argument because it seems that such models 
produce below the knee different spectra for p and other species. Note
 that even in usual acceleration models, collective effects 
can also produce such an effect~\cite{Ellison}.

\subsubsection{Data: behaviour at the knee}
The energy of the break
and slopes below and above the knee (denoted $\gamma_1$ and $\gamma_2$)
are in relative agreement between the various experiments (see 
Tab.~\ref{slopes}).
Direct measurements of fluxes and extraction from air showers experiments 
of proton and helium fluxes give a constant slope at least till 
$Z\times10^{15}$~eV~\cite{Apanasenko,Amenomori3,Clem}.
\begin{center}
\begin{table}[hbt!]
  \caption{Central values given by several experiments using several 
  Monte Carlo simulations for hadronic interactions.}
  \label{slopes}
  \begin{center}
  \begin{tabular}{|c|c|c|c|c|}\hline
  Experiment $[$ref$]$  & Simulation & $E_{\rm knee}$ &  $\gamma_1$ & 
  $\gamma_2$\\\hline\hline
 HEGRA~\cite{HEGRA} &{\sc qgsjet} & 4.0 PeV &2.72 & 3.22\\ 
 Tibet array~\cite{Amenomori1}& {\sc genas}& $\sim$ 2 PeV& 2.60& 3.00 \\ 
 EAS-TOP~\cite{Aglietta}&{\sc hdpm} &3 PeV& 2.76& 3.19 \\ 
 CASA-BLANCA~\cite{Fowler}& {\sc qgsjet}&$2-3$ PeV& 2.72& 2.95 \\ 
 KASCADE~\cite{Antoni1}& {\sc qgsjet}& 5.5 PeV& 2.77 & 3.11 \\ 
 KASCADE~\cite{Antoni1}& {\sc venus}& 4.5 PeV& 2.87 & 3.25\\\hline
  \end{tabular}
  \end{center}
\end{table}
\end{center}

However, 
measurements from $\langle\ln A\rangle$ have given contradictory 
conclusions (see {\em e.g.}~\cite{Glasmacher}): SAS~\cite{Shirasaki} gives
$\Delta \langle\ln A\rangle=+1.0$ between 2.5~--~6.3~PeV and constant 
above 6.3~PeV;
HEGRA~\cite{HEGRA} gives a composition consistent with direct measurements 
at 100~TeV and gradually becomes heavier around $\sim$~PeV; 
CASA-MIA~\cite{Glasmacher}
finds that $\langle\ln A\rangle$ increases in the range 100~TeV~--~10~PeV;
CASA-BLANCA~\cite{Fowler} observes a composition becoming lighter between 
1~--~3~PeV then heavier above 3~PeV. As pointed out in~\cite{Antoni1}, 
these apparent discrepancies could as well be related to the different 
interaction models used (see Tab.~\ref{slopes}
and Fig.~7 of~\cite{Fowler} where $\langle \ln A\rangle$
is displayed for four interaction models). At present {\sc qgsjet} and 
{\sc venus} seem to be favored~\cite{Fowler}, but it is clear that 
these Monte Carlo simulations are crucial to extract 
observables~\cite{Antoni1,Shirasaki2,Pop,Bakatanov}.

To sum up, from direct experiments and ground arrays,
we can confidently present the following 
results: there is no break in 
spectra before a few hundreds of TeV~\cite{Apanasenko} where usual cosmic 
ray acceleration is at work. Above a few PeV, the all-particle spectrum 
asymptotically reaches the slope $\sim 3.0$ (see Tab.~\ref{slopes}) 
extending up to a second break~\cite{AbuZayyad}. Data show a gradual 
steepening of the spectrum rather than a single kink, but still the 
steepening happens within about one decade of energy. Focusing on the average
logarithmic mass,  it is found that $\langle\ln A\rangle$ is about constant 
near the knee and then gradually increases above the break (a change 
from a heavy to a light composition is then observed in the energy region 
$5\times 10^{18}$~eV giving support to a different origin for these cosmic 
rays, {\em i.e.} extragalactic~\cite{AbuZayyad}).

Actually, Stronger constraints can be obtained by combining the all-particle spectrum and the
average mass: in~\cite{Bernlohr}, it is found that a simple model
with a universal slope and a break at a given rigidity can 
match either
the all-particle spectrum or  the average composition, but not 
both. Among four models tested, \cite{Shirasaki} find
that only one model (an adaptation of Biermann's one) is able to fit
both observables at the same time.

\subsubsection{Tentative model for the knee}

At this stage, the best way to explain the data is to generate this break 
at a given energy or rigidity or energy per nucleon with a change in 
slope $\Delta\gamma\sim-0.4$. Such a parameterization
is sufficient to see what happens in our diffusion model.
From what concerns the possibility of a different slope for p and other 
nuclei below the knee, we will set $\alpha_{\rm p}-\alpha_{\rm Z>1}= 0.05$,
althought greater values are possible ($\alpha_j$ denotes source spectra, 
whereas $\gamma_j$ corresponds to propagated/measured spectra).
Of course, the way we generate the break is to some extent
unphysical. It should be kept in mind that the propagated all-particle
spectrum and average logarithmic mass  -- which already present fine 
structures in this idealized case ({\em i.e.} sharp source spectral index 
break, see results in fig7 ) would be even more complicated in a more 
realistic model. Anyway, the exact form of this break is certainly 
very model-dependent, even for the particular case of transition between 
different accelerating sources, because of the variety of source that 
can be invoked.

    \subsection{Prescription for the propagation parameters}
    \label{Propagation_prescription} 
General solutions and discussions about diffusion models 
can be found in~\cite{Berezinskii}. The model we use
is cylindrically symmetric with two zones~-- thin disc and diffusive
halo~-- described at length in~\cite{PaperI} and also used in 
subsequent papers for primary~\cite{Barrau} and secondary 
antiprotons~\cite{Donato1}, and for $\beta$-radioactive 
nuclei~\cite{Donato2}.
The five parameters of this diffusion model are diffusive halo scale 
height $L$, rate of reacceleration related to the alfv\'enic speed 
$V_a$, galactic wind $V_c$, normalization $K_0$ and slope
$\delta$ of the diffusion coefficient (see Sec.~\ref{musea}).
A consistent range for the five parameters of our diffusion
model has been obtained~\cite{PaperI}, but there is a strong degeneracy 
between the various parameters derived.
Moreover, these parameters do not reproduce very well primary proton 
fluxes. A solution for this problem is seeked (Donato et al, 
in preparation) adding a sixth free parameter to the study, 
namely the source spectral index $\alpha_j$.

In this following, we focus on two specific parameters: the halo scale 
height $L$ and the slope of the diffusion coefficient $\delta$. The 
first one is related to geometrical effects
since a thin halo (say $\sim$ 3 kpc) would correspond to more ``local" 
 sources~-- in diffusion models, cosmic rays cannot come from regions 
 whose distance is greater than that of the closer edge --. The second one 
 is related to the
 source spectra $\alpha$, since the measured spectrum slopes at Earth 
 ($\gamma\approx 2.8$) are linked {\em via} the approximate relation 
 $\alpha\sim 
 (2.8-\delta)$, $\delta$ being the diffusion coefficient slope.
To extend these calculations above the knee, we have two possibilities: 
(i) keep the same parameters throughout the energy range, and explain 
the knee by a change in slope spectra (see previous section), or (ii) 
explain the knee 
by a change of diffusive regime. We previously noticed that
if the transition in diffusion regime is at a fixed rigidity, the situation
is thus strictly equivalent to a break in spectra at 
$E_{\rm knee}=Z$~eV~(see Sec.~\ref{gfdsa}). The transition proposed by 
Ptuskin and collaborators~\cite{PtuskinI} is rather smooth, but as it 
requires $\delta\sim 0.2$ that is excluded by Maurin et al.'s 
analysis~\cite{PaperI}, we will not consider anymore this possibility.
        \subsection{Radial distribution of sources $q(r)$}
  \label{radial_distrib}
Measurements of galactic $\gamma$ rays in the seventies
have raised the question of the radial distribution
of cosmic rays. This distribution is needed in order to evaluate
the resulting gamma emissivity at different galactocentric locations. 
The first distribution used was that of Kodaira (1974)~\cite{Kodaira}
following the radial distribution of supernovae which is also close
to that of pulsars. This is consistent with the present picture of 
cosmic rays where supernovae provide the energetic power and mechanism 
to accelerate nuclei. The description
of the galactocentric distribution has been improved
thanks to new observations of pulsars and supernovae. 
We take here the distribution of Case \& 
Bhattacharya~\cite{Case-Bhattacharya2}
which is an improvement of their earlier 
analysis~\cite{Case-Bhattacharya1}. 
Their last result provides a flatter distribution than the previous 
one, closer
to the distribution adopted by Strong \& Moskalenko~\cite{Strong0}. 
A third form that we finally retain is a constant radial distribution: 
this will serve as a reference to estimate the {\em pure} radial 
distribution effect on the average logarithmic mass. These three 
distributions are reproduced in 
Tab.~\ref{f_de_rho} and are presented in the left panel of 
Fig.~\ref{michel_elisa} 
with the effects of metallicity gradient (see next section).

        \subsection{Observed metallicity gradient}
  \label{gradientmetal}
The existence of radial metallicity gradients is
 now well established in spiral galaxies.
Early studies showed a gradient for O/H from observations of
ionized nebulae in galaxies like M33, M51, and M101, but later work
observed this trend in our Galaxy for many other abundances 
(see~\cite{Maciel1} for a review and Sec.~2 of~\cite{Gummersbach}). 
Several recent observations (see Tab.~1 of~\cite{Chiappini} for 
a compilation of results) lead to very similar conclusions for the 
metallicity gradient (see Table~\ref{Gradient}).

\begin{center}
\begin{table}[hbt!]
  \caption{Example of observed radial gradients (various elements)
  for several methods using samples distributed over the galactocentric
  distances $r_{GC}$.}
  \label{Gradient}
  \begin{center}
  \begin{tabular}{|l|c|c|c|c|}\hline
  Gradient      & Element & Range &  Tool & Ref.\\
  (dex kpc$^{-1}$)&    & ($r_{GC}$ in kpc)      &       &\\\hline\hline
  -0.02 / -0.05 & C, O,... Gd & 6-11 &  Cepheids & \cite{Andrievsky}\\
  -0.07 ($\pm$ 0.01)& C, O, Mg, Si & 6-18& Early B-type stars & 
  \cite{Rolleston}\\
  -0.055 ($\pm 0.015$)& O, Ne, S, Ar&4-10& Planetary nebulae& 
  \cite{Maciel}\\
  -0.09 & Fe & 8-16 & Open clusters & \cite{Carraro}\\
  -0.07 & N, O, S  & 0-12 & HII regions& \cite{Afflerbach}\\\hline
  \end{tabular}
  \end{center}
\end{table}
\end{center}

We choose the most recent study~\cite{Andrievsky} which takes 
into account more than 25 gradients corresponding to
species from carbon to gadolinium (see in particular their Fig.~6-9 
and Fig.~10). For all ions X (except He), the gradient is
about $d[{\rm X}/{\rm H}]/dr=-0.05$ dex kpc$^{-1}$.

        \subsection{Recapitulation of the model tested}
  \label{resume}
Let sum up the different models studied :
  \begin{enumerate}
    \item Source spectral index $\alpha$ -- see Sec.~\ref{biermanie}: 
    following our previous study~\cite{PaperI}, we set 
    $\alpha=2.8-\delta$.
      \begin{table}[hbt!]
        \caption{Models I and II used in this paper to describe the 
        slopes of cosmic ray sources, plus III and IV for the 
        description of the knee.}
        \label{IetII}
        \begin{center}
          \begin{tabular}{|c|c|}\hline
          Model I  & All species with the same $\alpha$\\
          Model II & H with $\alpha_{\rm p}$, all others with $\alpha$\\
          Model III & $\alpha$ below the knee and $\alpha_2=\alpha-0.4$ 
          above the knee\\
          Model IV & II below the knee, and $\alpha_2=\alpha-0.4$ for
          all species above the knee\\\hline
          \end{tabular}
        \end{center}
      \end{table}
    \item Propagation parameters -- see Sec.~\ref{Propagation_prescription}:
        three models with $\delta$
   (slope of the diffusion coefficient) equals to 0.46, 0.6 and 0.75
   at a fixed diffusive halo size $L=10$ kpc; one more with $L=3$  
   ($\delta=0.6$). Exact numbers for other parameters are unimportant,
   however we emphasize that for each value of $L$ 
   and $\delta$, the parameters $V_a$, $V_c$ and $K_0$ are set such as 
   to fit B/C data~\cite{PaperI}.
    \item Geometrical effects:
       \begin{itemize}
      \item Radial distribution $q(r)$ -- see Sec.~\ref{radial_distrib}: 
      three models.
        \begin{table}[hbt!]
            \caption{Models a,b and c used in this paper corresponding to 
            three possible radial distribution of sources.}
            \label{f_de_rho}
            \begin{center}
              \begin{tabular}{|c|cc|}\hline
              Model a  & $q(r)$= 1. &\\
              Model b & $\displaystyle q(r)=\left(\frac{r}{8.5}\right)^{2.0}
              \exp\left(-3.53\times\frac{(r-8.5)}{8.5}\right)$ & Case \&
              Bhattacharya~\cite{Case-Bhattacharya2} \\
              Model c & $\displaystyle q(r)=\left(\frac{r}{8.5}\right)^{0.5}
              \exp\left(-1.\times\frac{(r-8.5)}{8.5}\right)$ & Strong \&
              Moskalenko~\cite{Strong0}\\\hline
              \end{tabular}
            \end{center}
          \end{table}
      \item Metallicity gradient -- see Sec.~\ref{gradientmetal}:
      as one can see in formula~(\ref{qwerty}), there should be in the 
      above formula an additional multiplicative factor 
      $(q_{\rm X}/q_{\rm p})_{\odot}$ for all 
        \begin{table}[hbt!]
            \caption{Two models~-- one with and one without gradient~-- 
            used in this paper.}
            \label{0et1}
            \begin{center}
              \begin{tabular}{|c|c|}\hline
              Model 0 & No gradient\\
              Model 1 & Substitution $q(r) \rightarrow 10^{-0.05(r-8.)}
              \times q(r)$, except for H and He \\\hline
             \end{tabular}
            \end{center}
          \end{table}
  species $X$. It is implicitly taken into account since all fluxes
  are normalized to HEAO-3 data~\cite{heao-3} at 10.6~GeV/nuc, except for
  p and He that are respectively normalized to {\sc ams} proton data at 
  79.6~GeV~\cite{proton_ams} and helium data at 
  49.2~GeV/nuc~\cite{helium_ams}.
    \end{itemize}
  \end{enumerate}

Model Ia-0 will denote a model where all the sources have a fixed 
spectral index, the radial distribution is constant, and where
there is no gradient. Model IIb-1 will denote a model where 
source spectrum index of H is different from all others, where $q(r)$ is 
Case \& Bhattacharya's one (see above), and where we choose a composition 
gradient of -0.05 dex kpc$^{-1}$ for all species. 
We use this convention for all the figures proposed below.
The reference model from where other will be compared is Model Ia-0 with 
$\delta=0.6$, $L=10$ kpc and no break.
\section{Results and concluding remarks}
\label{ALL}
All outputs come from the diffusion model of~\cite{PaperI} with 
slight modifications of the inputs according to the above-prescriptions.

\subsection{Geometrical effects}
\begin{figure}[ht]
\includegraphics*[width=1.1\textwidth]{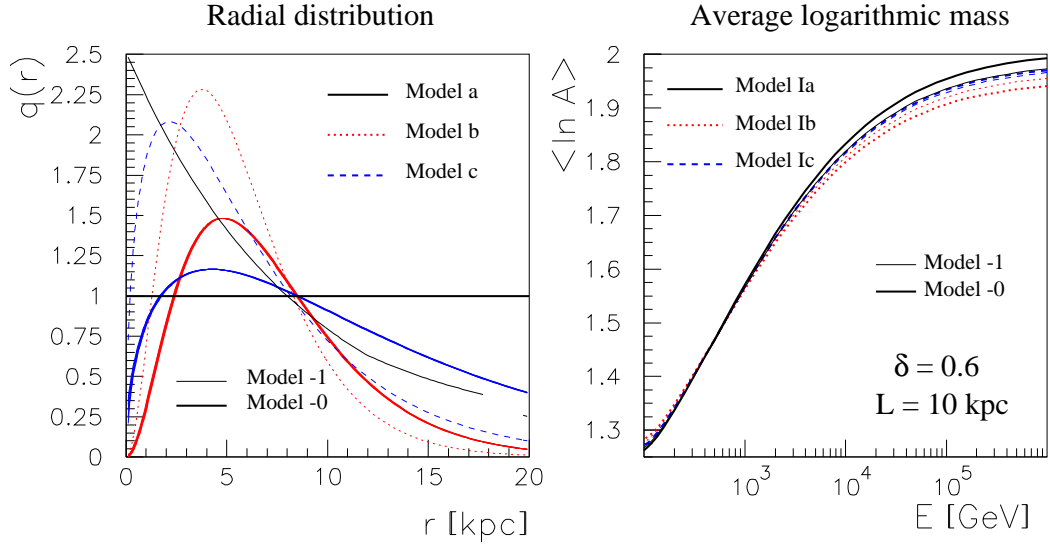}
\caption{Left panel: representation of the three radial distribution of 
sources: solid lines are for constant distribution (models~a), dotted 
lines are for Case \& Batthacharia (models~b) and dashed lines are for 
Strong \& Moskalenko (models~c); $q(r)$ is evaluated 
(see Tab.~\ref{0et1}) with (models~-1, thin line) or 
without (models~-0, thick line) metallicity gradient.
Right panel: average logarithmic mass (model I, $\delta=0.6$, $L=10$ kpc) 
for the six cases presented on left panel.}
\label{michel_elisa}
\end{figure}

Curve Ia-0 (Fig.~\ref{michel_elisa})
corresponds to a model
where all the species have the same spectral index, and where
no geometrical effect is allowed. The first strong conclusion
is that the {\em pure} propagation effect affects dramatically
the composition of cosmic rays. It is consistent with the conclusion of 
 Sec.~\ref{Propag_Effect}, namely that the evolution of the average mass is
closely connected to propagation properties. Furthermore, at sufficiently 
high energy, as expected, we reach the asymptotic regime where propagation 
ceases to affect $\langle \ln A\rangle$.

Other curves correspond to the evaluation of the {\em pure} 
geometrical effects. Both radial distribution (model a,b,c) and 
metallicity gradient (model 0,1) are separatly presented. 
In fact, comparing curve Ia-0 and Ia-1, we could conclude
prematuraly that metallicity effect plays a role in the evolution
of $\langle\ln A\rangle$. However, when 
the radial distribution of sources is correctly taken into account,
metallicity only has a little additional effect (compare curves 
Ib-0 and Ib-1). Finally, the impact of
metallicity is more or less pronounced depending of the 
distribution $q(r)$ chosen. Nevertheless, as was correctly 
guessed in Sec.~\ref{metalicite},
(i) metallicity effect is of little importance
and (ii) total geometrical effects correspond to an additional
change of at most 5\% compare to a non geometrical model.
Furthermore, Fig.~\ref{michel_elisa} corresponds to a halo scale 
height $L=10$~kpc, for which the effects are maximized. For $L=3$~kpc, 
these geometrical effects are completely negligible.

\subsection{Effects  of propagation and source spectra}
\begin{figure}[ht]
\includegraphics*[width=1.1\textwidth]{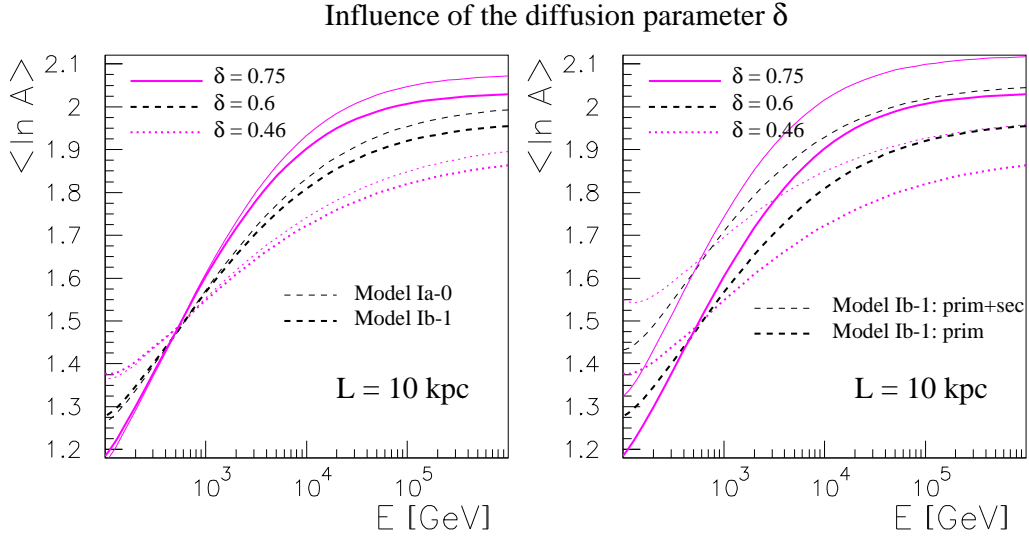}
\caption{Average logarithmic mass (model I, $L=10$ kpc) for three values
of the diffusion coefficient slope $\delta$. Left panel displays
these three values for two cases, (i) model Ia-0 (no geometrical 
effects, thin line) and (ii) model Ib-1 (Case \& Bhattacharia's 
q(r) with $\nabla [{\rm Fe}/{\rm H}]=-0.05$~dex~kpc$^{-1}$, 
thick line). Right panel shows for Model Ib-1 the same three 
$\delta$ values  but introducing the distinction between $\langle 
\ln A\rangle$ evaluated (i) with primary species only (thick line), 
(ii) with all primary and secondary nuclei (thin line).}
\label{arghhhh}
\end{figure}

Fig.~\ref{arghhhh} (left panel) shows 
that the cosmic ray composition is very sensitive to the diffusion 
parameters, the strongest
dependence being that of the diffusion power spectrum $\delta$ 
(the fixed point around 
500~GeV is just an artifact due to the normalization adopted).
 Note that the asymptotic diffusive regime is reached faster for 
 larger values of $\delta$.
 We checked that the influence of the parameter $L$ is minor 
 compared to $\delta$'s.
We note that secondary contribution (nuclei produced by spallation
 of the main primary species) 
is important (right panel). In particular,
when making the junction between direct measurements and ground arrays, 
these secondaries are almost never taken into account in the 
calculation of the average logarithmic mass, whereas they are 
implicitly counted in air showers 
data (their presence in the reconstructed quantity $\langle 
\ln A\rangle$ is more questionable~\cite{Antoni1}). 
This question is related to the ability of obtaining a confident 
normalization of $\langle\ln A\rangle$ with data from nuclear interaction 
models, {\em e.g.}~\cite{Fowler}.

\begin{figure}[ht]
\includegraphics*[width=1.1\textwidth]{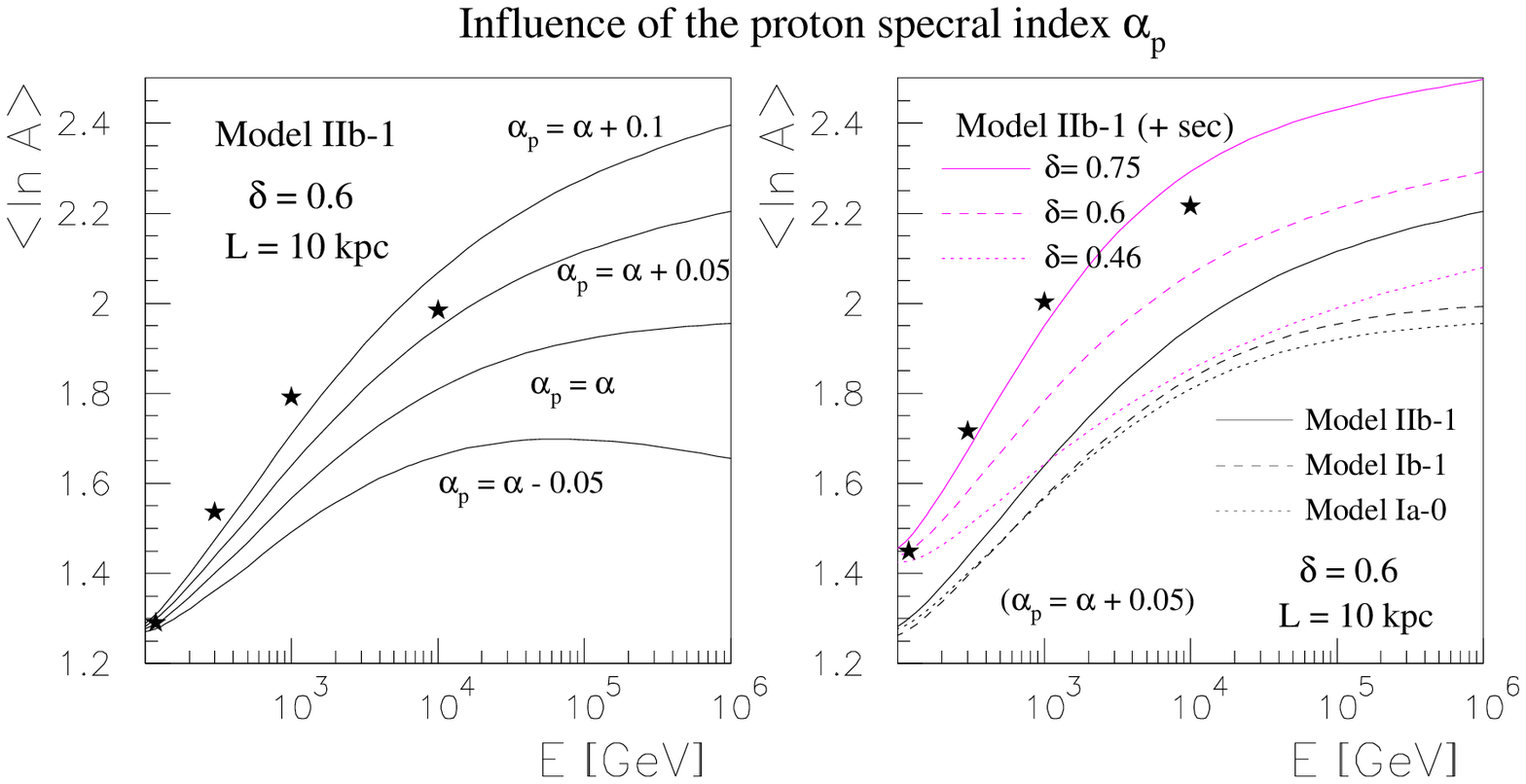}
\caption{Left panel: average logarithmic mass (model IIb-1, $\delta=0.6$, 
$L=10$ kpc) for various values of cosmic ray proton spectral index 
$\alpha_{\rm p}$ (see 
Tab.~\ref{IetII}).
Right panel: summary of the various effects investigated in
this paper for primary species, {\em i.e} {\em pure} propagation 
effect (Ia-0) + geometrical effect (Ib-1) + spectrum effect (IIb-1). 
The three upper curves correspond to three $\delta$ where secondary 
contributions are taken into account in Model IIb-1 
($\alpha_{\rm p}=\alpha+0.05$). Stars are values of $\langle\ln A\rangle$ 
measured in direct experiments (taken from Fig.~29 of~\cite{Apanasenko}) 
interpolated from their original normalization to our 100~GeV 
normalization value.}
\label{blurp}
\end{figure}

The impact of the spectral difference between protons 
(slope $\alpha_{\rm p}$) and other species (slope $\alpha_{\rm Z>1}\equiv 
\alpha$) is illustrated in Fig.~\ref{blurp} (left panel)
showing an increase of the mass composition with $\alpha_{\rm p}-\alpha$.
The right panel summarizes the astrophysical effects studied here
in the framework of a diffusion model: conclusions are similar to what
was drawn in Sec.~\ref{key}: pure propagation effects (Ia-0, 
dotted line) are mostly responsible for the increase of $\langle \ln 
A\rangle$ {\em vs} energy, geometrical effects (Ib-1, dashed line) 
are less significant, and source effects (IIb-1, solid line) turn 
from an almost asymptotically constant $\langle\ln A\rangle$ into 
a constant enhancement of the same quantity.
The three upper curves demonstrate importance of the diffusion 
power spectrum $\delta$ and emphasize the role of the secondaries 
in the normalization of the average logarithmic 
mass. We also display the average logarithmic mass as measured by 
experiments (stars). We renormalize to the observations at 100~GeV. 
We see that a difference $\alpha_{\rm p}-\alpha>0$ or/and large 
values of $\delta$ are preferred.

\subsection{Combination of all previous effects with a model for the knee}
\begin{figure}[ht]
\includegraphics*[width=1.1\textwidth]{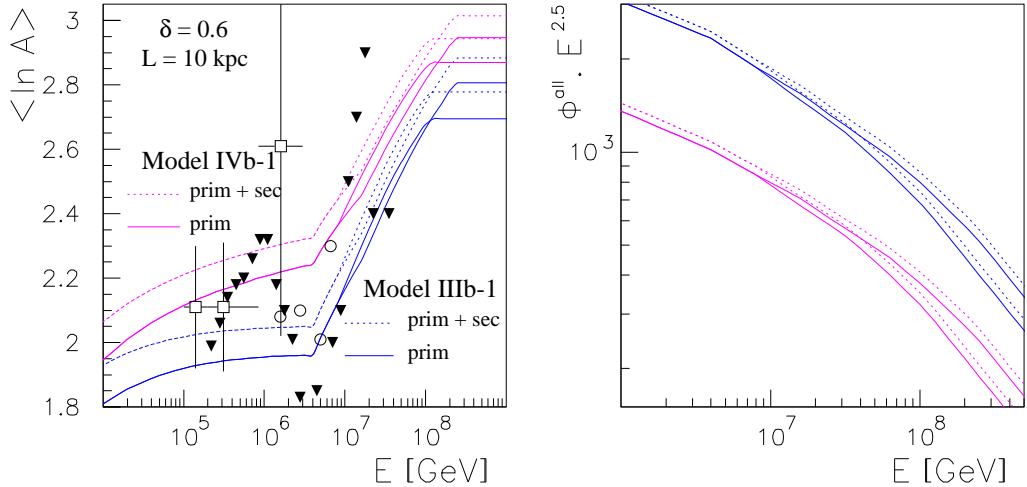}
\caption{Average logarithmic mass (left panel) and all-particle 
spectrum (right panel) for Model IIIb-1 and IVb-1 with a break either 
in rigidity ($R=4$~PeV) or in total energy per nucleus (E/nuc~$=4$~PeV). 
Solid lines correspond to primaries only, whereas dotted lines 
correspond to primaries plus secondaries. In the right panel, fluxes 
for Model IVb-1 have been divided by two in order to avoid the overlap 
with Model IIIb-1. For illustrative purpose, some data from ground 
arrays have been displayed: triangles are CASABLANCA's data with 
Monte-Carlo {\sc hdpm}~\cite{Fowler}, empty circles are from KASCADE 
collaboration with {\sc qgsjet} simulation~\cite{Antoni1} (empty 
squares are from JACEE direct experiment; RUNJOB data plus some 
JACEE data are lower than 1.8 and do not appear on the graph, 
see~\cite{Apanasenko}).
}
\label{blurpfinal}
\end{figure}

In Fig.~\ref{blurpfinal}, we generate the knee either 
with a break at a fixed rigidity, {\em i.e} 
$E_{\rm knee}=Z\times4$~PeV, or at a fixed total energy per nucleus 
{\em i.e} $E_{\rm knee}=A\times4$~PeV. The possibility of a break at 
a single energy is not considered because it exhibits a very sharp 
break in all-particle spectrum not present in data.
Fig.~\ref{blurpfinal} displays the resulting curves for the two models 
IIIb-1 and IVb-1 that only differ in their spectral indexes below 
the knee ($\alpha_{\rm p}-\alpha=0.05$ for 
model IVb-1 and $\alpha_{\rm p}-\alpha=0.$  for IIIb-1). Compared 
to the results of the two 
components model (see Sec.~\ref{gfdsa}), we remark both in $\langle 
\ln A\rangle$ and $\Phi^{\rm all}$ some additionnal bifurcations
generated by the helium component with a second even smoother 
transition provided by the CNO group. If these effects 
are not very relevant for the average mass composition, they hamper 
the interpretation of the transition from one regime to the other
in the all-particle spectrum data. Secondaries smooth even more these
transitions. They are also important for normalization of 
$\langle \ln A\rangle$. The two cases $E_{\rm knee}=Z\times4$~PeV and 
$E_{\rm knee}=A\times4$~PeV could be differentiated mostly through the 
all-particle spectrum. Finally, the bigger the difference 
between proton slope and other species before the knee, the smoother the 
bump in  $\langle \ln A\rangle$. 

Before concluding, 
we would like to make a brief comment on Swordy's model~\cite{Swordy2}: in 
the latter, an enhancement of light nuclei is predicted before the usual
enrichment in heavier nuclei at the knee (some data can support this
upturn, {\em e.g.} CASABLANCA's data, see Fig.~\ref{blurpfinal}). 
However, this requires a change in diffusion ({\em i.e} $\Delta\delta<0$
because $\Phi\propto R^{-(\gamma=\alpha+
\delta)}$) at an 
energy smaller than the knee's, so that 
it should produce a bump visible in all-particle spectrum: 
the larger $\Delta\delta$, the sharper the bump, and moreover a larger 
value of $\Delta\alpha$ is then necessary to reproduce the all-particle 
spectrum at the knee.
Thus, if position and sharpness of this change in diffusive regime (quite
constrained by all-particle spectra) is not theoritically excluded, 
it requires to be extremely fine tuned. Anyway, this model
along with our theoretical predictions cannot be tested due to the large 
data scattering (see Fig.~\ref{blurpfinal}).

The best clues up to now about the knee puzzle come from spectral analysis:
for example the KASCADE collaboration~\cite{Antoni2} find that the 
all-particle spectrum exhibits a knee  ($\Delta\gamma\sim 0.2-03$) 
around 4~PeV, but that this knee is seen only in their light ion 
subsample for which $\Delta\gamma\sim 0.5$. As regards the 
heavy ions, they find no changes in the region 1~--~10~PeV
but the slope below the knee is smaller than that of light component.
If this observation is confirmed, the average
logarithmic mass is likely to evolve as depicted in our Model IVb-1.
\subsection{Concluding remarks}
We have presented an analysis of the phenomena that affect the 
chemical cosmic ray composition up to highest ``Galactic" energies.
Several astrophysical effects
have been considered and geometrical effects have been found to play a 
minor role, while propagation effects (mostly selective 
destruction in flight of heavy nuclei) drive this evolution up to 
the knee where they cease to be effective.
A difference between
the source spectrum of protons and other ions lead to a constant 
enhancement of $\langle \ln A\rangle$ up to the knee and ceases if 
the slopes above the knee are similar for all species. In the framework 
of a simple break in rigidity (or total energy per nucleon), a bump 
in the chemical composition occurs at the knee, but the larger 
the spectral difference between protons and other species, the 
smoother the bump. The secondary species induce an enhancement 
in $\langle \ln A\rangle$ of about 15\%. As a by-product, our study 
validates the approach recently used in~\cite{Shirasaki}, {\em i.e}
we demonstrated that above PeV energy, a propagation model 
including only sources and diffusion is relevant. Finally, the main 
problem of the diffusive problem is the normalization of the fluxes. 
It could be improved thanks to a better determination of the 
propagation parameters with more precise low energy data.
\ack{We warmly thank Richard Taillet, Fiorenza Donato and Pierre Salati for 
very interesting discussions, and especially R. Taillet and F. Donato for 
useful comments on the paper. This work has been 
supported by the PICS number 1076 of INSU/CNRS.}

\end{document}